\documentstyle[twoside]{article}

\input FEYNMAN
\def\tr{{\rm tr}}

\makeatletter
\newdimen\normalarrayskip              
\newdimen\minarrayskip                 
\normalarrayskip\baselineskip
\minarrayskip\jot
\newif\ifold             \oldtrue            \def\new{\oldfalse}
\def\arraymode{\ifold\relax\else\displaystyle\fi} 
\def\@arrayskip{\ifold\baselineskip\z@\lineskip\z@
     \else
     \baselineskip\minarrayskip\lineskip2\minarrayskip\fi}
\def\@arrayclassz{\ifcase \@lastchclass \@acolampacol \or
\@ampacol \or \or \or \@addamp \or
   \@acolampacol \or \@firstampfalse \@acol \fi
\edef\@preamble{\@preamble
  \ifcase \@chnum
     \hfil$\relax\arraymode\@sharp$\hfil
     \or $\relax\arraymode\@sharp$\hfil
     \or \hfil$\relax\arraymode\@sharp$\fi}}
\def\@array[#1]#2{\setbox\@arstrutbox=\hbox{\vrule
     height\arraystretch \ht\strutbox
     depth\arraystretch \dp\strutbox
     width\z@}\@mkpream{#2}\edef\@preamble{\halign \noexpand\@halignto
\bgroup \tabskip\z@ \@arstrut \@preamble \tabskip\z@ \cr}%
\let\@startpbox\@@startpbox \let\@endpbox\@@endpbox
  \if #1t\vtop \else \if#1b\vbox \else \vcenter \fi\fi
  \bgroup \let\par\relax
  \let\@sharp##\let\protect\relax
  \@arrayskip\@preamble}
\makeatother

\catcode`\@=11
\long\def\@makefntext#1{
\protect\noindent \hbox to 3.2pt {\hskip-.9pt

$^{{\eightrm\@thefnmark}}$\hfil}#1\hfill}		

\def\thefootnote{\fnsymbol{footnote}}
\def\@makefnmark{\hbox to 0pt{$^{\@thefnmark}$\hss}}	

\def\ps@myheadings{\let\@mkboth\@gobbletwo
\def\@oddhead{\hbox{}
\rightmark\hfil\eightrm\thepage}

\def\@oddfoot{}\def\@evenhead{\eightrm\thepage\hfil
\leftmark\hbox{}}\def\@evenfoot{}
\def\sectionmark##1{}\def\subsectionmark##1{}}

\oddsidemargin=\evensidemargin
\renewcommand{\thefootnote}{\fnsymbol{footnote}}

\newcounter{sectionc}\newcounter{subsectionc}\newcounter{subsubsectionc}
\renewcommand{\section}[1] {\vspace{12pt}\addtocounter{sectionc}{1}

\setcounter{subsectionc}{0}\setcounter{subsubsectionc}{0}\noindent

	{\tenbf\thesectionc. #1}\par\vspace{5pt}}
\renewcommand{\subsection}[1] {\vspace{12pt}\addtocounter{subsectionc}{1}

	\setcounter{subsubsectionc}{0}\noindent

	{\bf\thesectionc.\thesubsectionc. {\kern1pt \bfit #1}}\par\vspace{5pt}}
\renewcommand{\subsubsection}[1] {\vspace{12pt}\addtocounter{subsubsectionc}{1}
	\noindent{\tenrm\thesectionc.\thesubsectionc.\thesubsubsectionc.
	{\kern1pt \tenit #1}}\par\vspace{5pt}}
\newcommand{\nonumsection}[1] {\vspace{12pt}\noindent{\tenbf #1}
	\par\vspace{5pt}}

\newcounter{appendixc}
\newcounter{subappendixc}[appendixc]
\newcounter{subsubappendixc}[subappendixc]
\renewcommand{\thesubappendixc}{\Alph{appendixc}.\arabic{subappendixc}}
\renewcommand{\thesubsubappendixc}
	{\Alph{appendixc}.\arabic{subappendixc}.\arabic{subsubappendixc}}

\renewcommand{\appendix}[1] {\vspace{12pt}
        \refstepcounter{appendixc}
        \setcounter{figure}{0}
        \setcounter{table}{0}
        \setcounter{lemma}{0}
        \setcounter{theorem}{0}
        \setcounter{corollary}{0}
        \setcounter{definition}{0}
        \setcounter{equation}{0}
        \renewcommand{\thefigure}{\Alph{appendixc}.\arabic{figure}}
        \renewcommand{\thetable}{\Alph{appendixc}.\arabic{table}}
        \renewcommand{\theappendixc}{\Alph{appendixc}}
        \renewcommand{\thelemma}{\Alph{appendixc}.\arabic{lemma}}
        \renewcommand{\thetheorem}{\Alph{appendixc}.\arabic{theorem}}
        \renewcommand{\thedefinition}{\Alph{appendixc}.\arabic{definition}}
        \renewcommand{\thecorollary}{\Alph{appendixc}.\arabic{corollary}}
        \renewcommand{\theequation}{\Alph{appendixc}.\arabic{equation}}
        \noindent{\tenbf Appendix \theappendixc #1}\par\vspace{5pt}}
\newcommand{\subappendix}[1] {\vspace{12pt}
        \refstepcounter{subappendixc}
        \noindent{\bf Appendix \thesubappendixc. {\kern1pt \bfit #1}}
	\par\vspace{5pt}}
\newcommand{\subsubappendix}[1] {\vspace{12pt}
        \refstepcounter{subsubappendixc}
        \noindent{\rm Appendix \thesubsubappendixc. {\kern1pt \tenit #1}}
	\par\vspace{5pt}}

\newcommand{\textlineskip}{\baselineskip=13pt}
\newcommand{\smalllineskip}{\baselineskip=10pt}

\def\eightcirc{
\begin{picture}(0,0)
\put(4.4,1.8){\circle{8.5}}
\end{picture}}
\def\eightcopyright{\eightcirc\kern2.7pt\hbox{\eightrm c}}

\newcommand{\copyrightheading}[1]
	{\vspace*{-2.5cm}\smalllineskip{\flushleft
	{\footnotesize International Journal of Modern Physics A, #1}\\
	{\footnotesize $\copyright$\, World Scientific Publishing
	 Company}\\
	 }}

\def\abstracts#1#2#3{{
	\centering{\begin{minipage}{4.5in}\baselineskip=10pt\footnotesize
	\parindent=0pt #1\par

	\parindent=15pt #2\par
	\parindent=15pt #3
	\end{minipage}}\par}}

\newcommand{\bibit}{\nineit}

\renewenvironment{thebibliography}[1]
	{\frenchspacing
	 \ninerm\baselineskip=11pt
	 \begin{list}{\arabic{enumi}.}
	{\usecounter{enumi}\setlength{\parsep}{0pt}
	 \setlength{\leftmargin 12.7pt}{\rightmargin 0pt} 
	 \setlength{\itemsep}{0pt} \settowidth
	{\labelwidth}{#1.}\sloppy}}{\end{list}}

\newcounter{itemlistc}
\newcounter{romanlistc}
\newcounter{alphlistc}
\newcounter{arabiclistc}

\newcommand{\fcaption}[1]{
        \refstepcounter{figure}
        \setbox\@tempboxa = \hbox{\footnotesize Fig.~\thefigure. #1}
        \ifdim \wd\@tempboxa > 5in
           {\begin{center}
        \parbox{5in}{\footnotesize\smalllineskip Fig.~\thefigure. #1}
            \end{center}}
        \else
             {\begin{center}
             {\footnotesize Fig.~\thefigure. #1}
              \end{center}}
        \fi}

\newcommand{\tcaption}[1]{
        \refstepcounter{table}
        \setbox\@tempboxa = \hbox{\footnotesize Table~\thetable. #1}
        \ifdim \wd\@tempboxa > 5in
           {\begin{center}
        \parbox{5in}{\footnotesize\smalllineskip Table~\thetable. #1}
            \end{center}}
        \else
             {\begin{center}
             {\footnotesize Table~\thetable. #1}
              \end{center}}
        \fi}

\def\@citex[#1]#2{\if@filesw\immediate\write\@auxout
	{\string\citation{#2}}\fi
\def\@citea{}\@cite{\@for\@citeb:=#2\do
	{\@citea\def\@citea{,}\@ifundefined
	{b@\@citeb}{{\bf ?}\@warning
	{Citation `\@citeb' on page \thepage \space undefined}}
	{\csname b@\@citeb\endcsname}}}{#1}}

\newif\if@cghi
\def\cite{\@cghitrue\@ifnextchar [{\@tempswatrue
	\@citex}{\@tempswafalse\@citex[]}}
\def\citelow{\@cghifalse\@ifnextchar [{\@tempswatrue
	\@citex}{\@tempswafalse\@citex[]}}
\def\@cite#1#2{{$\null^{#1}$\if@tempswa\typeout
	{IJCGA warning: optional citation argument

	ignored: `#2'} \fi}}

\def\pmb#1{\setbox0=\hbox{#1}
	\kern-.025em\copy0\kern-\wd0
	\kern.05em\copy0\kern-\wd0
	\kern-.025em\raise.0433em\box0}

\def\fnt#1#2{\footnotetext{\kern-.3em
	{$^{\mbox{\scriptsize #1}}$}{\footnotesize\smalllineskip\baselineskip=10pt
#2}}}

\def\fpage#1{\begingroup
\voffset=.3in
\thispagestyle{empty}\begin{table}[b]\centerline{\footnotesize #1}
	\end{table}\endgroup}

\def\runninghead#1#2{\pagestyle{myheadings}
\markboth{{\protect\footnotesize\it{\quad #1}}\hfill}
{\hfill{\protect\footnotesize\it{#2\quad}}}}
\font\tenrm=cmr10
\font\tenit=cmti10

\font\tenbf=cmbx10
\font\bfit=cmbxti10 at 10pt
\font\ninerm=cmr9
\font\nineit=cmti9

\font\eightrm=cmr8

\def\qed{\hbox{${\vcenter{\vbox{			
   \hrule height 0.4pt\hbox{\vrule width 0.4pt height 6pt
   \kern5pt\vrule width 0.4pt}\hrule height 0.4pt}}}$}}

\renewcommand{\thefootnote}{\fnsymbol{footnote}}	

\begin{document}

\runninghead{G. Amelino-Camelia, I. I. Kogan and R. J. Szabo}
{Gravitational Dressing of Aharonov-Bohm Amplitudes}

\normalsize\textlineskip
\thispagestyle{empty}
\setcounter{page}{1}

\copyrightheading{}			

\vspace*{0.88truein}

\fpage{1}
\begin{flushright}

OUTP-96-60P 

hep-th/9610057

October 1996

\end{flushright}
\centerline{\bf GRAVITATIONAL DRESSING OF}
\vspace*{0.035truein}
\centerline{\bf AHARONOV-BOHM AMPLITUDES\footnote{Presented 
by G. Amelino-Camelia as plenary talk
at the Workshop on Low Dimensional Field
Theory, Telluride, CO, 5-17 Aug 1996. 
To be published in the proceedings.}}
\vspace*{0.37truein}
\centerline{\footnotesize Giovanni AMELINO-CAMELIA,
Ian I. KOGAN and Richard J. SZABO}
\vspace*{0.015truein}
\centerline{\footnotesize\it Department of Theoretical Physics, University of
Oxford}
\baselineskip=10pt
\centerline{\footnotesize\it 1 Keble Road, Oxford OX1 3NP, U.K.}
\vspace*{0.225truein}

\vspace*{0.21truein}
\abstracts{\baselineskip=10pt We
investigate Aharonov-Bohm scattering
in a theory in which
charged bosonic matter fields
are coupled to
topologically massive electrodynamics
and topologically massive gravity.
We demonstrate that,
at one-loop order,
the transmuted spins in this theory
are related to the ones of ordinary Chern-Simons
gauge theory in the same way that
the Knizhnik-Polyakov-Zamolodchikov formula
relates
the Liouville-dressed
conformal weights of primary operators
to the bare weights
in two-dimensional conformal field theories.
We remark on the implications of this connection between
two-dimensional conformal field theories
and three-dimensional gauge and gravity theories
for a topological membrane reformulation
of strings.
We also discuss
some features of the gravitational analog of
the Aharonov-Bohm effect.}{}{}

\textheight=7.8truein
\setcounter{footnote}{0}
\renewcommand{\thefootnote}{\alph{footnote}}

\textlineskip			
\vspace*{12pt}			
\noindent
The Aharonov-Bohm effect\cite{ab} is one of the most extensively studied
problems in
planar physics.
Originally,
the primary motivation for its investigation
was the fact that this effect
requires a physical, rather than mathematical,
interpretation of the electromagnetic gauge potential.
A second surge
of interest in
the subject
was sparked by the realization
that flux-charge composites ({\it anyons})
exhibit fractional statistics through the Aharonov-Bohm effect\cite{wil}.
Such composites can be described by
coupling ordinary particles to a Chern-Simons gauge field,
so that the Aharonov-Bohm effect is equivalently encoded in
the scattering of a charged
particle from a flux tube, the scattering of two
flux-charge composites ({\it two-anyon scattering}),
and the scattering of two
particles coupled to a Chern-Simons gauge field
({\it Aharonov-Bohm scattering}).
Substantial work has also been devoted to
the development of successful perturbative approaches
within non-relativistic\cite{abdelta}
and relativistic\cite{kogmor,pertrel} formalisms, in light of the
intriguing observation\cite{corinaldesi}
that the perturbative expansion
of the expression originally obtained by
Aharonov and Bohm cannot be obtained
within the ordinary
Born approximation.
Recently, we have also advocated\cite{a-cks}
the study of the
(generalized)
Aharonov-Bohm effect
as a tool in the investigation
of the relations
between certain three-dimensional gauge and gravity theories
and some associated two-dimensional conformal field theories.
Here we shall discuss
Aharonov-Bohm scattering
in a theory in which
charged bosonic
matter fields are coupled to
topologically massive electrodynamics
and topologically massive gravity,
and use it to probe the relation between
(two-dimensional) Liouville theory
and topologically massive gravity.

The action of
topologically massive gauge theory\cite{djt} is
\begin{equation}
S_{TMGT}^{[{\cal G}]} = -\frac{1}{2e^2} \int_{\cal
M}d^3x~\tr~\sqrt{g}~g^{\mu\lambda}g^{\nu\rho}F_{\mu\nu}F_{\lambda\rho}+ S_{CS}
\label{tmgt}\end{equation}
where
\begin{equation} S_{CS} = \int_{\cal M} {k\over4\pi}~\tr\left(A\wedge dA +
{2\over3}A\wedge A\wedge A\right)
\label{csaction}\end{equation}
is the topological, parity-violating Chern-Simons action, the three-dimensional
space-time 
manifold $\cal M$ has metric $g_{\mu\nu}$ of Minkowski signature, and
$A=A_\mu(x)dx^\mu=A_\mu^a(x)T^adx^\mu$ is a gauge connection of a trivial
vector bundle over $\cal M$ with $T^a$ the anti-Hermitian generators of the
compact gauge group ${\cal G}$.
The first term in (\ref{tmgt}) is the usual Yang-Mills
kinetic term for the gauge fields with $F=dA+[A,A]/2$ the curvature of $A$.
Unlike the pure Chern-Simons theory (\ref{csaction}), the action (\ref{tmgt})
does not define a topological field theory because the $F^{2}$-term depends
explicitly on the three-dimensional metric $g_{\mu\nu}$ and there are
propagating degrees of freedom (massive vector bosons) with topological mass
\begin{equation}
M=ke^2/4\pi
\end{equation}
so that the Hilbert space of this quantum field theory is infinite-dimensional.

The action of topologically massive gravity\cite{djt} is
\begin{equation}
S_{TMG}=\frac{k'}{8\pi}\int_{\cal M}\left(\omega^a\wedge
d\omega^a+\frac{2}{3}\epsilon^{abc}\omega^a\wedge\omega^b\wedge\omega^c\right)
+S_E
\label{tmg}\end{equation}
where
\begin{equation}
S_{E}=\kappa\int_{\cal M}e^a\wedge R^a
\label{e}\end{equation}
is the three-dimensional Einstein gravity action,
$e^a=e_\mu^adx^\mu$ are the dreibein fields which generate local
$SO(2,1)$ rotations of the frame bundle of $\cal M$ (and are related to
the metric of $\cal M$ by $\eta_{ab}e_\mu^ae_\nu^b=g_{\mu\nu}$,
with $\eta^{ab}$ the local flat Minkowski
metric of the tangent bundle $T{\cal M}$),
$\kappa$ is the Planck mass, and
\begin{equation}
R^a=R^a_{\mu\nu}dx^\mu\wedge
dx^\nu=d\omega^a+\epsilon^{abc}\omega^b\wedge\omega^c
\label{curvom}\end{equation}
is the curvature of the spin-connection
$\omega^a=\epsilon^{abc}\omega^{bc}=\epsilon^{abc}\omega_\mu^{bc}dx^\mu$
of the frame bundle of $\cal M$.
The dreibein and spin-connection fields are related by the constraint
\begin{equation}
\nabla e^a=de^a+2\epsilon^{abc}\omega^b\wedge e^c=0
\label{LCconstr}\end{equation}
which ensures that $\omega^a$ is the Levi-Civita spin-connection for
$g_{\mu\nu}$ (this means that we are working in the `minimal' formalism of
general relativity). Pure three-dimensional
Einstein gravity ($k' \! = \! 0$ in (\ref{tmg})) can be viewed as a topological
Chern-Simons theory with gauge group $ISO(2,1)$\cite{Town},
but the full gravity theory (\ref{tmg})
is not topological. The graviton degrees of
freedom propagate with topological mass
\begin{equation}
\mu=8\pi\kappa/k'
\label{gravmass}\end{equation}
The first term in (\ref{tmg}) can
be regarded as a Chern-Simons action for an
$SO(2,1)$ gauge theory with connection $\omega$.

We shall be concerned with
Aharonov-Bohm scattering
in the theory in which
charged, massive scalar fields are minimally coupled to
topologically massive electrodynamics
and topologically massive gravity,
{\it i.e.} the theory described by the action
\begin{equation}
S= S_{TMGT}^{[U(1)]} + S_{TMG} + \int_{\cal M} d^3x~ \sqrt{g} \, \left(
g^{\mu\nu}[(\partial_\mu-iA_\mu)\phi]^*[(\partial_\nu-iA_\nu)\phi]
-m^2~\phi^*\phi \right)
\label{matgravaction}\end{equation}
where $g=\det[g_{\mu\nu}]$ and $g^{\mu\nu}=\eta^{ab}e^\mu_ae^\nu_b$. The
Aharonov-Bohm amplitude
resulting from the theory (\ref{matgravaction})
can be expected, based on the expected relation\footnote{There are
several results\cite{carlip3,gravrel} suggesting
that topologically massive gravity might be related to
Liouville theory.
In particular, some evidence in support of
this relation is provided in Ref.\cite{carlip3} 
at the level of a (formal) 
path integral analysis.
One can interpret the result we report in the present paper,
as an indication that a suitable quantum measure 
does exist such that the results of the analysis of Ref.\cite{carlip3} 
hold at the full quantum level.} between (dressing by) 
topologically massive gravity
and (dressing by) Liouville theory,
to be related to the ordinary
Aharonov-Bohm amplitude
in a simple way, dictated
by the Knizhnik-Polyakov-Zamolodchikov (KPZ)
scaling relations\cite{pol2} for primary fields
of Liouville theory
\begin{equation}
{\hat \Delta}-{\hat \Delta}_0=\frac{{\hat \Delta}(1-{\hat \Delta})}{c+2}
\label{KPZtransfLiouville}\end{equation}
where $c$ is the central charge of the
$SL(2, R)$ current algebra,
${\hat \Delta_0}$ is the
bare conformal dimension of the primary field,
and ${\hat \Delta}$ is its Liouville-dressed conformal dimension.

As discussed in detail in Ref.\cite{a-cks}
and references therein,
the formal relation between
topologically massive gravity
and Liouville theory identifies
the conformal dimensions of primary fields in the latter
with the transmuted spins that charged
particles acquire due to their interaction with the
former.\footnote{In general,
the transmuted spins that
(three-dimensional) charged
particles acquire due to their interaction with
a Chern-Simons gauge field
are equal\cite{a-cks} to
the conformal dimensions of the primary fields of
a corresponding
(two-dimensional) Wess-Zumino-Novikov-Witten model.}
Moreover, the central charge $c$ of Liouville theory
is related to the gravitational Chern-Simons
coefficient in (\ref{tmg}) by $c=-k'-4$.
Accordingly, from (\ref{KPZtransfLiouville})
it would follow that
the transmuted spins $\Delta$
of the theory (\ref{matgravaction})
should be related
to the ``bare" transmuted spins $\Delta_0$
(the ones that charged
particles acquire as a result of their interaction with the
ordinary Chern-Simons gauge field)
according to
\begin{equation}
\Delta-\Delta_0=\frac{\Delta(\Delta - 1)}{k'+2}
\label{KPZtransf}\end{equation}
In turn this equation encodes the relation between
the Aharonov-Bohm amplitude
resulting from the theory (\ref{matgravaction})
and the ordinary
Aharonov-Bohm amplitude.
In fact, the transmuted spins
completely specify these amplitudes.
To review this, we
minimally couple charged particles in an irreducible unitary
representation $R({\cal G})$ of the gauge group ${\cal G}$
to the Chern-Simons gauge field
$A$ with a conserved current $J^\mu=J_a^\mu R^a$.
Then the invariant amplitude
for the scattering of two charged particles
of initial momenta $p_1$ and $p_2$
represented by the current $J$ in the infrared limit $M\to\infty$
is\cite{a-cks} (Fig.~1)
\begin{equation}\new{\begin{array}{lll}
{\cal A}(p_1,p_2;q)&\equiv&\lim_{M\to\infty} i~\tr~J^\mu(2p_1-q)
G_{\mu\nu}(q)J^\nu(2p_2+q)\\&=&-\frac{16\pi
i}{k}\dim({\cal G})T_R({\cal G})f_{\cal
G}(k)\frac{\epsilon_{\mu\nu\lambda}p_1^\mu p_2^\nu
q^\lambda}{q^2}\end{array}}
\label{abrelgen}\end{equation}
where
\begin{equation}
G_{\mu\nu}(p)=\left\langle
A_\mu(p)A_\nu(-p)\right\rangle_A= - i e^2 \left(
\frac{p^2g_{\mu\nu}^\perp(p)+iM
\epsilon_{\mu\nu\lambda}p^\lambda}{p^2(p^2-M^2)}\right)
\label{tmgaugeprop}\end{equation}
is the momentum space bare gluon propagator of the topologically massive gauge
theory (\ref{tmgt}) in the transverse covariant Landau gauge, with
$g_{\mu\nu}^\perp(p)=g_{\mu\nu}-p_\mu p_\nu/p^2$ the transverse projection
operator on the momentum space of vectors,
$q$ is the momentum transfer, $T_R({\cal G})$ is the quadratic
Casimir of ${\cal G}$
in the representation $R({\cal G})$, 
and $f_{\cal G}(k)=\sum_{n\geq0}f_n/k^n$
is a function whose coefficients $f_n$
(which can be computed perturbatively
order by order in the Chern-Simons
coupling constant $1/k$)
depend only on invariants of the gauge group ${\cal G}$.
In the center of
momentum frame the amplitude (\ref{abrelgen}) is none other than
the Aharonov-Bohm amplitude
for the scattering of a charge of strength $\sqrt{T_R({\cal G})f_{\cal G}(k)}$
off of a flux
of strength $(4\pi/k)\dim({\cal G})\sqrt{T_R({\cal G})f_{\cal G}(k)}$. This is
the standard argument
for the appearance of induced fractional spin and statistics
perturbatively in
a Chern-Simons gauge theory and it leads to the spin factor
({\it transmuted spin})
\begin{equation}
\Delta_{\cal G}(k)=\frac{T_R({\cal G})}{k}f_{\cal G}(k)
\label{spin}\end{equation}
which measures the anomalous change of phase in the
Aharonov-Bohm wavefunction
under adiabatical rotation of one charged particle
about another in the gauge
theory (\ref{csaction}).\cite{kogmor}

\begin{figure}[htbp]
\begin{center}
\begin{picture}(70000,8000)
\small
\put(19000,3500){\makebox(0,0){$q$}}
\put(12000,1000){\makebox(0,0){$p_2$}}
\put(25000,1000){\makebox(0,0){$p_2+q$}}
\put(12000,6000){\makebox(0,0){$p_1$}}
\put(25000,6000){\makebox(0,0){$p_1-q$}}
\drawline\fermion[\E\REG](18000,1000)[5000]
\drawline\fermion[\W\REG](18000,1000)[5000]
\drawline\photon[\N\REG](18000,1000)[5]
\put(18000,1000){\circle*{500}}
\put(18000,6000){\circle*{500}}
\drawline\fermion[\E\REG](\photonbackx,\photonbacky)[5000]
\drawline\fermion[\W\REG](\photonbackx,\photonbacky)[5000]
\end{picture}
\end{center}
\vspace*{13pt}
\fcaption{\baselineskip=10pt The scattering amplitude for two charged particles
in topologically massive Yang-Mills theory. Here $p_1,p_2$ denote the incoming
particle momenta and $q$ is the momentum transfer. Straight lines denote
external charged particles, wavy lines depict gluons, and the solid circles
represent the minimal coupling of the particle current $J^\mu$ to the gluon
field.}
\end{figure}

Eqs.~(\ref{abrelgen}) and (\ref{spin})
completely specify the relation between
Aharonov-Bohm amplitudes and transmuted spins,
and therefore in the following we report results on
the Aharonov-Bohm amplitude
by simply reporting the corresponding transmuted spin.
Since we intend to discuss the
Aharonov-Bohm amplitude
in the theory (\ref{matgravaction})
up to one-loop order by
viewing the topologically massive gravity theory
as a quantum field theory on
a flat space,
we shift the dreibein fields as
\begin{equation}
e_\mu^a \to e_\mu^a+ \delta_\mu^a
\label{dreishift}\end{equation}
and expand the various metric factors in
(\ref{matgravaction}). The one-loop action is then
\begin{equation}\new{\begin{array}{lll}
S^{(1)}[A_\mu,\phi,e_\mu^a]&=&\int
d^3x~\left\{-\frac{1}{4e^2}F^2+|(\partial-iA)\phi|^2-m^2|\phi|^2+
\frac{k}{8\pi}AdA\right.\\&
&~~~-\frac{1}{4e^2}\left(\eta^{\mu\lambda}\eta^{\nu\rho}
e^\alpha_\alpha-2\eta^{\mu\lambda}\eta^{\rho
a}e_a^\nu-2\eta^{\nu\rho}\eta^{\lambda
a}e_a^\mu-\frac{3}{2}\eta^{\mu\lambda}\eta^{\nu\rho}e^\alpha_\sigma
e^\sigma_\alpha\right.\\&
&~~~+\eta^{\mu\lambda}e^\nu_ae^\rho_a+\eta^{\nu\rho}e^\mu_ae^
\lambda_a+4\eta^{\lambda a}\eta^{\rho b}e^\mu_ae^\nu_b\\&
&\left.~~~-2\eta^{\mu\lambda}\eta^{\rho
a}e^\nu_ae_\alpha^\alpha-2\eta^{\nu\rho}\eta^{\lambda
a}e^\mu_ae_\alpha^\alpha+\frac{1}{2}\eta^{\mu\lambda}\eta^{\nu\rho}
e^\alpha_\alpha e^\sigma_\sigma\right)F_{\mu\nu}F_{\lambda\rho}\\&
&~~~+\left(\eta^{\mu\nu}
e^\lambda_\lambda+2\eta^{\nu a}e_a^\mu-\eta^{\mu\nu}e_\lambda^\rho
e_\rho^\lambda-e^\mu_ae^\nu_a\right.\\& &\left.~~~+2\eta^{\nu a}e_a^\mu
e_\lambda^\lambda+\frac{1}{2}\eta^{\mu\nu}e^\lambda_\lambda
e^\rho_\rho\right)[(\partial_\mu-iA_\mu)\phi]^*[(\partial_\nu-iA_\nu)
\phi]\\& &\left.~~~-\frac{m^2}{4}\left(2e^\mu_\mu+e^\mu_\mu
e^\nu_\nu-3e^\mu_\nu e^\nu_\mu\right)\phi^*\phi\right\}+S_{TMG}[e_\mu^a\to
e_\mu^a+\delta_\mu^a]\end{array}}
\label{1loopgravmat}\end{equation}
{}From (\ref{1loopgravmat}) the Feynman rules associated with each of the
graviton-matter interactions relevant for the total one-loop scattering
amplitude can be easily written down\cite{a-cks}. In particular, the bare
propagator for the dreibein field in momentum space and
in the Landau gauge is given by the Deser-Yang graviton
propagator\cite{a-cks,desyang}
\begin{equation}\new{\begin{array}{l}
D_{\mu\nu}^{ij}(p)=\langle e_\mu^i(p)e_\nu^j(-p)\rangle_e
\\=\frac{i}{\kappa}\left(\frac{\mu^2}{2p^2(p^2-\mu^2)}
\left\{\left(\frac{p^2}{\mu^2}-2\right)\eta_{\mu\nu}^\perp(p)\eta^{\perp
ij}(p)+\delta_\mu^{\perp i}(p)\delta_\nu^{\perp j}(p)+\delta_\mu^{\perp
j}(p)\delta_\nu^{\perp i}(p)\right\}\right.\\
{}~~~\left.+\frac{i\mu}{4}\frac{p^\lambda}{p^2(p^2-\mu^2)}\left\{\epsilon_{\mu~
\lambda}^{~i}\delta_\nu^{\perp
j}(p)+\epsilon_{\mu~\lambda}^{~j}\delta_\nu^{\perp
i}(p)+\epsilon_{\nu~\lambda}^{~i}\delta_\mu^{\perp
j}(p)+\epsilon_{\nu~\lambda}^{~j}\delta_\mu^{\perp
i}(p)\right\}\right)\end{array}}
\label{gravprop}\end{equation}
The Aharonov-Bohm amplitude for the scattering of charged particles
in the
topologically massive quantum field theories arises
from the imaginary,
parity-odd part of the propagator for the mediating bosons
which has a singular
pole term at zero momentum.
We want to evaluate
the one-graviton corrections
to the ordinary Aharonov-Bohm amplitude,
{\it i.e.} we wish to determine, to order $1/k'$, the transformation of the
bare
transmuted spin, which in the pure, topologically massive electromagnetic case
is given by\cite{pertrel,gaugeren}
\begin{equation}
\Delta_0=1/k
\label{spin0}\end{equation}
and is exact at tree level\cite{gaugeren},
due to the gravitational dressing.

The one-loop gravitational dressing of the parity-odd part of the topologically
massive photon propagator vanishes identically in the infrared regime of the
theory.\cite{lerdavan} Consequently,
the gravitational contributions to the vacuum polarization do not affect the
Aharonov-Bohm interaction at one-loop order.
Furthermore, a careful analysis\cite{a-cks}
of the gravitational dressing of the
topologically massive electrodynamics
in (\ref{matgravaction}) reveals that the
standard set of Ward-Takahashi identities
for pure quantum electrodynamics holds as well
for the gravitational renormalizations, and this implies\cite{a-cks}
that the total one-loop
gravitational renormalization of the
Aharonov-Bohm amplitude comes from the
contributions of the ladder diagrams depicted in Fig.~2.
Actually, in Fig.~2
we have depicted only the topologically
inequivalent Feynman diagrams. There
are 9 ladder graphs in total.
The ``triangle" diagram in Fig.~2 has a
counterpart corresponding
to $p_1\leftrightarrow p_2$, $q\to-q$ and the ``box"
graph has a partner
corresponding to $p_2\to-(p_2+q)$. Also, with the exception
of the last ``circle" diagram, each diagram has a partner
associated with interchanging the photon
and graviton lines (equivalently the interchange of
initial and final external particles in the diagrams).

\begin{figure}[htbp]
\begin{center}
\begin{picture}(40000,7000)
\small
\drawline\fermion[\E\REG](5000,1000)[7000]
\drawline\fermion[\E\REG](5000,5000)[7000]
\drawline\photon[\N\REG](7000,1000)[4]
\drawline\gluon[\N\REG](10000,1200)[3]
\drawline\fermion[\N\REG](10000,1000)[200]
\drawline\fermion[\N\REG](10000,4500)[500]
\drawline\fermion[\E\REG](14000,1000)[8000]
\drawline\fermion[\E\REG](14000,4800)[8000]
\drawline\photon[\SE\REG](18000,4800)[6]
\drawline\gluon[\SW\FLIPPED](18000,4800)[3]
\drawline\fermion[\E\REG](24000,1000)[6000]
\drawline\fermion[\E\REG](24000,4800)[6000]
\drawline\photon[\NW\REG](27000,1000)[3]
\drawline\gluon[\NE\FLIPPED](27000,1000)[1]
\drawline\photon[\SW\FLIPPED](27000,4800)[3]
\drawline\gluon[\SE\REG](27000,4800)[1]
\end{picture}
\end{center}
\vspace*{13pt}
\fcaption{\baselineskip=10pt The topologically inequivalent one-loop
gravitational ladder (one-photon and one-graviton exchange) diagrams. The
spiral lines depict the dreibein fields $e_\mu^a$.}
\end{figure}

The computation of the ladder amplitudes depicted in Fig.~2 are rather
involved, even in the infrared regime $M,\mu\to\infty$ of the quantum field
theory.
Each of the Feynman integrals for these ladder amplitudes converges in the
infrared limit $e^2,\kappa\to\infty$ and the final results for the functions
$f(k)$ defined in (\ref{abrelgen}) are independent of any mass scale of the
model. The leading order $1/k'$
coefficients of the functions associated with
the three amplitudes in Fig.~2 are\cite{a-cks}
\begin{equation}
f_0^{(\Box)}=-\frac{14+13\pi^2}{2k'}~~~~~,~~~~~
f_0^{(\bigtriangleup)}=
\frac{29+26\pi^2}{4k'}~~~~~,~~~~~f_0^{(\bigcirc)}=0
\label{ladderspins}\end{equation}
Taking into account the remaining six ladder diagrams which come from permuting
the external particle lines in the Feynman graphs in Fig.~2, the total
conformal dimension up to one-loop order is then
\begin{equation}
\Delta_{\rm
grav}^{(1)}=\Delta_0+(4f_0^{(\Box)}
+4f_0^{(\bigtriangleup)}+f_0^{(\bigcirc)})/k
=1/k+1/kk'
\label{KPZ1loop}\end{equation}
which, taking into account (\ref{spin0}) and
selecting the branch of (\ref{KPZtransf}) with
$\Delta(\Delta_0=0)=0$,
agrees with the leading orders of the (iterative)
large-$k'$ expansion
of the KPZ
formula (\ref{KPZtransf}).\footnote{From the point 
of view of the two-dimensional Liouville
theory, there is no immediate reason 
to choose the $\Delta(\Delta_0=0)=0$ branch for the solutions of
the KPZ scaling relations; however, 
our approach selects this 
branch automatically. It would be interesting to give a
three-dimensional  interpretation of the other branch of the scaling relation
(\ref{KPZtransf}).}
Having established this relation
at least to one-loop order of perturbation theory
provides further evidence in support of the expected 
relation\cite{carlip3,gravrel}
between topologically massive gravity and
Liouville theory,
and
is encouraging for the topological membrane approach to string
theory\cite{kogan1}, in which
the string world-sheet
is filled in and
viewed as the boundary of a three-manifold.

We conclude with a few comments concerning the
possibility of a gravitational analogue
of the Aharonov-Bohm effect, which could result from
the parity-odd structure in (\ref{gravprop})
that has a simple pole at $p^2 \! = \! 0$.
Recent work\cite{gravanyon},
based on a non-perturbative
analysis within an abelian, linearized
approximation to the action,
has advocated this effect based on
the observation that
spinless, non-dynamical point particles in the
presence of a topologically massive gravity field acquire an induced spin
$k'/32\pi^2$ under adiabatical rotation. However, our perturbative
large-$k'$ analysis does not (and could not) expose 
this non-perturbative phenomenon. 
In particular,
the ``tree-level gravitational
Aharonov-Bohm amplitude'', associated to the
imaginary part of the Feynman diagram in Fig.~1
with the photon line replaced
by a graviton line, vanishes identically for all ranges of momenta
as a result of the property\cite{a-cks}
\begin{equation}
{\cal T}^{({\rm grav})}(p_1,p_2;q)^{\rm odd}
=i{\cal E}_i^\mu(p_1,p_1-q){\cal E}_j^\nu(p_2,p_2+q)
D_{\mu\nu}^{ij}(q)^{\rm odd}\equiv0
\label{treegrav}\end{equation}
where we have considered
only the parity-odd ``$\epsilon$-terms'' of the full
graviton propagator (\ref{gravprop}), and ${\cal E}_i^\mu(p,p';q=p-p')$
is the bare meson-meson-graviton vertex
function determined from (\ref{1loopgravmat}). This feature
is also true of the higher-loop amplitudes involving only graviton lines, 
and is a result of the index contractions 
that occur in the integrands of the
Feynman integrals. Moreover,
our one-loop computations show that there
is no induced gravitational Aharonov-Bohm effect
from the interactions of the
topologically massive graviton
field with the meson and photon fields, {\it i.e.}
we find that, for all ranges of momenta, the
integrands of the corresponding Feynman integrals
which come from contracting
the parity-even part of the photon propagator
with the parity-odd part of the graviton
propagator vanish identically.
Since the initial spin of the charged particles
is zero if only gravitational interactions
(no gauge fields)
are turned on, this perturbative
result is consistent with the KPZ formula (\ref{KPZtransf}) for the branch
that has $\Delta(\Delta_0=0)=0$, in which each term
is at least of order $\Delta_0$,
{\it i.e.} there are no terms at any given
order of the $1/k'$ expansion which are not
accompanied by factors of $\Delta_0$.
It is only for particles with non-zero bare
spin, such as fermions or vector bosons,
that one could find perturbatively non-vanishing pure
gravitational corrections.
In these cases, the non-zero diagrams come from the coupling of the spinning
fields to the spin-connection $\omega_\mu^a$ itself.

The framework of our analysis is different 
from the one of the
analysis of the gravitational Aharonov-Bohm
effect reported in Ref.~\cite{gravanyon},
in which
a mass source was coupled
to gravity to yield a static space-time 
with conical singularities.\cite{star}
Such a singularity acts as a source for an Aharonov-Bohm type
amplitude,
as illustrated in Ref.~\cite{djth} 
within a study
of gravitational Aharonov-Bohm
scattering. 
It would be interesting to connect the low-energy
amplitude obtained above with the conical structure of space-time, but this
relation is still not understood. We saw above that the gravitational
interaction changes the original Aharonov-Bohm amplitude in a flat space-time.
{}From our calculation it is clear that this effect is a result of
energy-momentum being induced by a charge which was a source for those
``gravitons" and yielded an extra contribution to the Aharonov-Bohm
amplitude. It would be interesting to find a suitable solution of the field
equations for topologically massive gravity interacting with topologically
massive gauge theory (see Ref.~\cite{kogan2} for a discussion of this and some
exact solutions in the pure Einstein theory). Because our result 
is valid only for 
distances much larger than the inverse graviton mass, {\it i.e.} 
the region of space-time where this renormalized amplitude is derived is
the region of pure Einstein gravity, the space-time behaves like a cone. The
source of mass inside the cone is the area filled by the electric and magnetic
fields from the topologically massive gauge theory source, and as a result the
deficit angle of the cone is proportional to the bare weight $\Delta_0$. Thus
it appears that the renormalized weight $\Delta$ includes this cone angle.
However, the problem now is that 
in order to find an appropriate solution of the field
equations one has to match the outer region of the cone with the inner region
where the full topologically massive gauge and gravity theory must be taken
into account. Some interesting results along these lines have been obtained
recently in Ref.~\cite{clement}, but the necessary solution has yet to be
found.

\nonumsection{Acknowledgements}
\noindent
G.A.-C. acknowledges very useful discussions with
A. Cappelli, S. Deser, R. Jackiw, E. Mottola, M. Ortiz,
and G. Zemba.
The work of G.A.-C. was supported in part by PPARC.
The work of R.J.S. was supported in
part by the Natural Sciences and Engineering Research Council of Canada.


\nonumsection{References}


\begin{thebibliography}{000}
\newcommand{\NPB}[1]{{\bibit Nucl.~Phys.}~{\bf B#1}}
\newcommand{\Ann}[1]{{\bibit Ann.~Phys.}~{\bf #1}}
\newcommand{\CMP}[1]{{\bibit Commun.~Math.~Phys.}~{\bf #1}}
\newcommand{\PLB}[1]{{\bibit Phys.~Lett.}~{\bf B#1}}
\newcommand{\PRL}[1]{{\bibit Phys.~Rev.~Lett.}~{\bf #1}}
\newcommand{\PTP}[1]{{\bibit Progr.~Theor.~Phys.}~{\bf #1}}
\newcommand{\MPLA}[1]{{\bibit Mod.~Phys.~Lett.}~{\bf A#1}}
\newcommand{\IJMP}[1]{{\bibit Int.~J.~Mod.~Phys.}~{\bf A#1}}
\newcommand{\IJMPB}[1]{{\bibit Intern.~J.~Mod.~Phys.}~{\bf B#1}}
\newcommand{\CQG}[1]{{\bibit Class.~Quant.~Grav.}~{\bf #1}}
\newcommand{\PR}[1]{{\bibit Phys.~Rev.}~{\bf #1}}
\newcommand{\PRD}[1]{{\bibit Phys.~Rev.}~{\bf D#1}}
\newcommand{\PRB}[1]{{\bibit Phys.~Rev.}~{\bf B#1}}
\newcommand{\JMP}[1]{{\bibit J.~Math.~Phys.}~{\bf #1}}

\bibitem{ab} Y. Aharonov and D. Bohm, \PR{115}, 485 (1959).

\bibitem{wil} F. Wilczek, \PRL{48}, 1144 (1982).

\bibitem{abdelta} O. Bergman and G. Lozano, \Ann{229}, 416 (1994); D. Bak and
O. Bergman, \PRD{51}, 1994 (1995); G. Amelino-Camelia and D. Bak, \PLB{343},
231 (1995); C. R. Hagen, \PRD{52}, 2466 (1995);
P. Giacconi, F. Maltoni and R. Soldati, {\it ibid.} {\bf D53}, 952 (1996); G.
Amelino-Camelia, \PLB{326}, 282 (1994);
C. Manuel and R. Tarrach, {\it ibid.} {\bf B328}, 113 (1994);
S. Ouvry, \PRD{50}, 5296 (1995);
G. Amelino-Camelia, {\it ibid.} {\bf D51}, 2000 (1995);
A. Comtet, S. Mashkevich, 
and S. Ouvry, {\it ibid.} {\bf D52}, 2594 (1995);
S.-J. Kim and C. Lee, hep-th/9606054.

\bibitem{kogmor} I. I. Kogan and A. Yu. Morozov, {\bibit Sov. Phys. JETP} {\bf
61}, 1 (1985); M. I. Dobroliubov, D. Eliezer, I. I. Kogan, G. W. Semenoff and
R. J. Szabo, \MPLA{8}, 2177 (1993).

\bibitem{pertrel} I. I. Kogan, {\it JETP Lett.} {\bf 49}, 225 (1989); A.
Groshev and E. R. Poppitz, {\it Phys. Lett.} {\bf B235}, 336 (1990); H. O.
Girotti, M. Gomes and A. J. da Silva, {\it ibid.} {\bf B274}, 357 (1992); I. I.
Kogan and G. W. Semenoff, \NPB{368}, 718 (1992); R. J. Szabo, I. I. Kogan and
G. W. Semenoff, {\it ibid.} {\bf B392}, 700 (1993); Y. Georgelin and J. C.
Mallet, Phys. Rev. {\bf D50}, 6610 (1994).

\bibitem{corinaldesi} E. L. Feinberg,
{\bibit Usp.~Fiz.~Nauk.}~{\bf 78}, 53 (1962)
[{\bibit Sov.~Phys.~Usp.}~{\bf 5}, 753 (1963)];
E. Corinaldesi and F. Rafeli,
{\bibit Am.~J.~Phys.}~{\bf 46}, 1185 (1978).

\bibitem{a-cks} G. Amelino-Camelia, I. I. Kogan and R. J. Szabo,
hep-th/9607037, {\bibit Nucl.~Phys.}~{\bf B} (1996), in press.

\bibitem{djt} S. Deser, R. Jackiw and S. Templeton, \Ann{\bf 140}, 372 (1982).

\bibitem{Town} A. Achucaro and P. K. Townsend, \PLB{180}, 89 (1986); E. Witten,
\NPB{311} (1988), 46; {\it ibid.} {\bf B323}, 113 (1989); S. Carlip, {\it
ibid.} {\bf B324}, 106 (1989).

\bibitem{carlip3} S. Carlip,  \NPB{362}, 111 (1991).

\bibitem{gravrel} S. Carlip and I. I. Kogan, \PRL{64}, 148 (1990); {\it ibid.}
{\bf 67}, 3647 (1991); \MPLA{6}, 171 (1991); S. Carlip, \PRD{45}, 3584 (1992);
I. I.  Kogan, \PLB{256}, 369 (1991);
I. I. Kogan, \NPB{375}, 362 (1992); 
M. C. Ashworth, hep-th/9510192.

\bibitem{pol2} A. M. Polyakov, \MPLA{2}, 893 (1987); V. G. Knizhnik, A. M.
Polyakov and A. B. Zamolodchikov, {\it ibid.} {\bf A3}, 819 (1988); F. David,
{\it ibid.} {\bf A3}, 1651 (1988); J. Distler and H. Kawai, \NPB{321}, 509
(1989).

\bibitem{desyang} S. Deser and Z. Yang, \CQG{7}, 1603 (1990).

\bibitem{gaugeren} I. Affleck, J. A. Harvey and E. Witten, \NPB{206}, 413
(1982); R. Jackiw and A. N. Redlich, \PRL{50}, 555 (1983); A. J. Niemi and G.
W. Semenoff, {\bibit ibid.} {\bf 51}, 2077 (1983); S. Coleman and B. Hill,
\PLB{159}, 184 (1985); R. D. Pisarski and S. Rao, \PRD{32}, 2081 (1985); G. W.
Semenoff, P. Sodano and Y.-S. Wu, \PRL{62}, 715 (1989); W. Chen, G. W. Semenoff
and Y.-S. Wu, \PRD{46}, 5521 (1992); G. Giavarini, C. P. Martin and F. Ruiz
Ruiz, \NPB{381}, 222 (1992).

\bibitem{lerdavan} A. Lerda and P. van Nieuwenhuizen, \PRL{62}, 1217 (1989); C.
R. Hagen and S. Ramaswamy, {\bibit ibid.} {\bf 63}, 213 (1989); C. Pinheiro, G.
O. Pires and F. A. B. Rabelo de Carvalho, {\bibit Europhys. Lett.} {\bf 25},
329 (1994).

\bibitem{kogan1} I. I. Kogan, \PLB{231}, 377 (1989).

\bibitem{gravanyon} S. Deser, \PRL{64}, 611 (1990); \CQG{9}, 61 (1992); S.
Deser and J. G. McCarthy, \NPB{344}, 747 (1990); M. Reuter, \PRD{44}, 1132
(1991).

\bibitem{star} A. Staruszkiewicz, {\it Acta Phys. Pol.} {\bf 24}, 734 (1963).

\bibitem{djth} S. Deser, R. Jackiw and G. 't Hooft, \Ann{152}, 220 (1984).

\bibitem{kogan2} I. I. Kogan, \MPLA{7}, 2341 (1992).

\bibitem{clement} G. Cl\'ement, \PLB{367}, 70 (1996); K. Ait Moussa and G.
Cl\'ement, \CQG{13}, 2319 (1996).

\end{thebibliography}
\end{document}